\documentclass[a4paper,11pt]{article}
\pdfoutput=1 
\usepackage{latexsym}
\usepackage{amsmath,amssymb,amsfonts,wasysym}
\usepackage{graphics}
\usepackage{xcolor}
\usepackage{float}
\usepackage{subfigure}
\usepackage{longtable}
\usepackage{verbatim}
\usepackage{cite}
\usepackage{jheppub} 
\usepackage[utf8]{inputenc}
\numberwithin{equation}{section}

\def\({\left(}
\def\){\right)}

\def\be{\begin{equation}}
\def\ee{\end{equation}}

\input{tcilatex}

\title{Radiatively generated hierarchy of lepton and quark masses}
\author{A. E. C\'arcamo Hern\'andez, }
\emailAdd{antonio.carcamo@usm.cl}
\author{Sergey Kovalenko, }
\emailAdd{sergey.kovalenko@usm.cl}
\author{Ivan Schmidt}
\emailAdd{ivan.schmidt@usm.cl}
\affiliation{Centro Cient\'{\i}ficoTecnol\'ogico de Valpara\'{\i}so-CCTVal,\\
Universidad T\'ecnica Federico Santa Mar\'{\i}a,\\
Casilla 110-V, Valpara\'{\i}so, Chile}
\date{\today }

\abstract{We propose a model for radiatively generating the hierarchy of  the Standard Model (SM) fermion masses: tree-level top quark mass;
1-loop bottom, charm, tau and muon masses; 2-loop masses for the light up, down and strange quarks as well as for
the electron; and 4-loop masses for the light active neutrinos. Our
model is based on a softly-broken $S_{3}\times Z_{2}$ discrete symmetry. Its 
scalar sector consists only of one SM Higgs doublet and three electrically
neutral SM-singlet scalars. We do not need to invoke either electrically
charged scalar fields, or an extra $SU_{2L}$ scalar doublet, or the spontaneous 
breaking of the discrete group, which are typical for other radiative models in the literature. The model features a viable scalar dark matter candidate.}

\begin{document}
\maketitle

\section{Introduction.}
\label{intro} 
Despite its 
striking consistency with experimental data, the Standard Model (SM) is
unable to explain several fundamental issues, such as the number of fermion
generations, the observed pattern of fermion masses and mixings, etc. To
provide an explanation of the \mbox{SM-fermion} mass hierarchy, several extensions
of the SM with radiative seesaw mechanisms have been 
constructed in the literature \cite%
{Ma:1988fp,Ma:1989ys,Dong:2006gx,Hernandez:2013mcf,Hernandez:2013dta,Campos:2014lla,Boucenna:2014ela,Okada:2015bxa,Wang:2015saa,Arbelaez:2016mhg,Nomura:2016emz,Kownacki:2016hpm,Nomura:2016ezz,Hernandez:2015hrt}. 

Here we propose an economical radiative  model, 
which explains the fermion mass hierarchy by a sequential loop
suppression, so that the masses are generated according to: \vspace{-4mm} 
\begin{eqnarray}
t\mbox{-quark} &\rightarrow &\mbox{{\it tree-level mass} from}\ \ \ \ 
\overline{q}_{jL}\widetilde{\phi }u_{3R},  \label{eq:level-1} \\
b,c,\ \tau ,\mu &\rightarrow &\mbox{\it 1-loop mass;}\ \mbox{tree-level}\label{eq:level-2} \\
&&\hspace{20mm}\mbox{suppressed by a {\it symmetry}}.  \notag \\
s,u,d,\ e &\rightarrow &\mbox{\it 2-loop mass;}\ \mbox{tree-level \& 1-loop}\label{eq:level-3}
\\
&&\hspace{20mm}\mbox{suppressed by a {\it symmetry}}.  \notag \\
\nu _{i} &\rightarrow &\mbox{\it 4-loop mass;}\ 
\mbox{tree-level \& lower
loops} \label{eq:level-4}\\
&&\hspace{20mm}\mbox{suppressed by a {\it symmetry}}. \notag \\[-3mm]
&&  \notag
\end{eqnarray}%
The operator in (\ref{eq:level-1}) is the SM Yukawa coupling. 
This scenario implies that, 
for the mass matrices $M_{U,D}$ of 
up and down quarks, $M_{l,\nu }$, of charged leptons and light active
neutrinos 
\begin{eqnarray}
M_{U} &=&\left( 
\begin{array}{ccc}
\widetilde{\varepsilon }_{11}^{\left( u\right) } & \varepsilon _{12}^{\left(
u\right) } & \kappa_{13}^{\left( u\right) } \\ 
\widetilde{\varepsilon }_{12}^{\left( u\right) } & \varepsilon _{22}^{\left(
u\right) } & \kappa_{23}^{\left( u\right) } \\ 
\widetilde{\varepsilon }_{13}^{\left( u\right) } & \varepsilon _{32}^{\left(
u\right) } & \kappa_{33}^{\left( u\right) }%
\end{array}%
\right) \frac{v}{\sqrt{2}},  \label{Mu} \\
M_{D} &=&\left( 
\begin{array}{ccc}
\widetilde{\varepsilon }_{11}^{\left( d\right) } & \widetilde{\varepsilon }%
_{12}^{\left( d\right) } & \varepsilon _{13}^{\left( d\right) } \\ 
\widetilde{\varepsilon }_{21}^{\left( d\right) } & \widetilde{\varepsilon }%
_{22}^{\left( d\right) } & \varepsilon _{23}^{\left( d\right) } \\ 
\widetilde{\varepsilon }_{31}^{\left( d\right) } & \widetilde{\varepsilon }%
_{32}^{\left( d\right) } & \varepsilon _{33}^{\left( d\right) }%
\end{array}%
\right) \frac{v}{\sqrt{2}},  \label{Md}
\end{eqnarray}%
\begin{eqnarray}
M_{l} &=&\left( 
\begin{array}{ccc}
\widetilde{\varepsilon }_{11}^{\left( l\right) } & \varepsilon _{12}^{\left(
l\right) } & \varepsilon _{13}^{\left( l\right) } \\ 
\widetilde{\varepsilon }_{21}^{\left( l\right) } & \varepsilon _{22}^{\left(
l\right) } & \varepsilon _{23}^{\left( l\right) } \\ 
\widetilde{\varepsilon }_{31}^{\left( l\right) } & \varepsilon _{32}^{\left(
l\right) } & \varepsilon _{33}^{\left( l\right) }%
\end{array}%
\right) \frac{v}{\sqrt{2}},  \label{Ml} \\
M_{\nu } &=&\left( 
\begin{array}{ccc}
\varepsilon _{11}^{\left( \nu \right) } & \varepsilon _{12}^{\left( \nu
\right) } & \varepsilon _{13}^{\left( \nu \right) } \\ 
\varepsilon _{12}^{\left( \nu \right) } & \varepsilon _{22}^{\left( \nu
\right) } & \varepsilon _{23}^{\left( \nu \right) } \\ 
\varepsilon _{13}^{\left( \nu \right) } & \varepsilon _{23}^{\left( \nu
\right) } & \varepsilon _{33}^{\left( \nu \right) }%
\end{array}%
\right) \frac{v^{2}}{\sqrt{2}\, \Lambda},  \label{Mnu2} 
\end{eqnarray}%
their entries are generated at different loop-levels: 
\begin{eqnarray}\label{eq:MassHierarch-1}
\kappa_{j3}^{\left(u\right)} &\rightarrow& \mbox{tree-level}\\
\label{eq:MassHierarch-2}
\varepsilon _{j2}^{\left( u\right)}, \varepsilon _{j3}^{\left( d\right) }, 
\varepsilon _{j2}^{\left(l\right)}, \varepsilon _{j3}^{\left(l\right)}  &\rightarrow& \mbox{1-loop-level}\\
\widetilde{\varepsilon }_{j1}^{\left( u\right) }, \widetilde{\varepsilon }_{j1}^{\left( d\right) }, 
\widetilde{\varepsilon }_{j2}^{\left( d\right) }, \widetilde{\varepsilon }_{j1}^{\left( l\right) }  &\rightarrow& \mbox{2-loop-level}\\
\label{eq:MassHierarch-4}
\varepsilon _{jk}^{\left( \nu \right) }  &\rightarrow& \mbox{4-loop-level},
\end{eqnarray}
%
where $ j,k=1,2,3$.
Since the SM fermion masses appear after the
electroweak symmetry breaking, the mass matrices are proportional to the VEV 
$v=\langle \phi ^{0}\rangle $ of the SM Higgs, $\phi$, or $v^{2}$ in the case of the
neutrinos. The latter is the  generic consequence of the fact that with the SM-fermion content,
the only possibility for the neutrino mass is the Majorana
option, described by the Weinberg operator $LL^{c}\phi \phi $. The mass
parameter $\Lambda$ in Eq.~(\ref{Mnu2}) is the scale of this operator, which
will be introduced in what follows.\newline

\section{The model setup}
\label{Model}
First we specify our model, allowing for the implementation of
the above described setup and then discuss the justification of its
structure. 
Let us stress right from the beginning that we do not pretend to explain the fine details of 
the mass matrices (\ref{Mu})-(\ref{Mnu2}) and therefore nor to predict the experimental values of the quark and lepton masses and mixings. Our goal is more moderate: to provide a mechanism underlying 
the hierarchy (\ref{eq:MassHierarch-1})-(\ref{eq:MassHierarch-4}). 
Towards this end we extend the SM gauge group $G_{SM}=SU_{3c}\times SU_{2L}\times U_{1Y}$
with the discrete symmetry factor\\[-3mm]
\begin{equation}
\mathcal{G}=G_{SM}\times S_{3}\times Z_{2},
\label{eq:symm-1}
\end{equation}%
\\[-5mm]
which is selected to be the least extra symmetry necessary for the
suppression pattern in Eqs. (\ref{eq:level-1})-(\ref{eq:level-4}). More
comments will be given below. The field content of the model consists of the SM fields augmented only with
$SU_{2L}$ singlets.

The scalar sector consists of the SM doublet Higgs $\phi $ and three
SM-singlets $\sigma _{1}$, $\sigma _{2}$, $\eta $, with the $S_{3}\otimes
Z_{2}$ assignments: 
\begin{equation}
\phi \sim \left( \mathbf{1},1\right) ,\hspace{1cm}\sigma=\left(\sigma_1,\sigma_2\right) \sim \left( \mathbf{%
2},1\right) ,\hspace{1cm}\eta \sim \left( \mathbf{1},-1\right) ,
\end{equation}
Let us note that aforementioned scalar content is the minimal required to implement the radiative mechanism of the SM fermion mass hierarchy generation  (\ref{eq:level-1})-(\ref{eq:level-4}).
%
%
The $S_3$ doublet SM-singlet scalar field $\sigma$ allows us  to implement only the 1-loop stage of this mechanism 
for the  charged fermion masses (\ref{eq:level-2}). On the other hand, both the $S_3$ trivial singlet SM-singlet scalar field $\eta$ and the $S_3$ doublet SM-singlet scalar $\sigma$ are needed to implement 2-loop level masses for the light up, down and strange quarks as well as for the electron.
It is worth noting that for the 4-loop 
light
active neutrino mass generation (\ref{eq:level-4}), we do not need charged
scalar fields, which are typical in the proposal of Ref.~\cite{Nomura:2016fzs}.

The fermion sector of the SM is extended with $SU_{2L}$ singlet exotic
quarks $T$, $\tilde{T}$, $B$, $\tilde{B}$ and singlet leptons $E$, $\tilde{E}$, $\nu _{s}$ ($s=1,2$)
with electric charges $Q(T)=Q(\tilde{T})=2/3$, $Q(B)=Q(\tilde{B})=-1/3$, 
$Q(E)=-1$. The $S_{3}\times Z_{2}$ assignments of the fermion sector are 
%
\begin{eqnarray}
u_{1R} &\sim &\left( \mathbf{1}^{\prime },-1\right) ,\hspace{0.5cm}%
u_{2R}\sim \left( \mathbf{1}^{\prime },1\right) ,\hspace{0.5cm}u_{3R}\sim
\left( \mathbf{1},1\right) ,  \notag  
\label{eq:Assign-1} \\
d_{1R} &\sim &\left( \mathbf{1}^{\prime },-1\right) ,\hspace{0.5cm}%
d_{2R}\sim \left( \mathbf{1}^{\prime },-1\right) ,\hspace{0.5cm}d_{3R}\sim
\left( \mathbf{1}^{\prime },1\right) ,  \notag \\
l_{1R} &\sim &\left( \mathbf{1}^{\prime },-1\right) ,\hspace{0.5cm}%
l_{2R}\sim \left( \mathbf{1}^{\prime },1\right) ,\hspace{0.5cm}l_{3R}\sim
\left( \mathbf{1}^{\prime },1\right) ,  \notag \\
q_{jL} &\sim &\left( \mathbf{1},1\right) ,\hspace{0.5cm}l_{jL}\sim \left( 
\mathbf{1},1\right) ,\hspace{0.5cm}j=1,2,3,  \notag \\
T_{L} &=&\left( T_{1L},T_{2L}\right) \sim \left( \mathbf{2},1\right) ,%
\hspace{0.2cm}T_{R}=\left( T_{1R},T_{2R}\right) \sim \left( \mathbf{2}%
,1\right) ,  \notag \\
\widetilde{T}_{L} &=&\left( \widetilde{T}_{1L},\widetilde{T}_{2L}\right)
\sim \left( \mathbf{2},1\right) ,\hspace{0.2cm}\widetilde{T}_{R}=\left( 
\widetilde{T}_{1R},\widetilde{T}_{2R}\right) \sim \left( \mathbf{2}%
,-1\right) ,  \notag \\
B_{L} &=&\left( B_{1L},B_{2L}\right) \sim \left( \mathbf{2},1\right) ,%
\hspace{0.2cm}B_{R}=\left( B_{1R},B_{2R}\right) \sim \left( \mathbf{2}%
,1\right) ,  \notag \\
\hspace{0.2cm}\widetilde{E}_{L} &=&\left( \widetilde{E}_{1L},\widetilde{E}%
_{2L}\right) \sim \left( \mathbf{2},1\right) ,\hspace{0.2cm}\widetilde{E}%
_{R}=\left( \widetilde{E}_{1R},\widetilde{E}_{2R}\right) \sim \left( \mathbf{%
2}-1\right) ,  \notag \\
\widetilde{B}_{L}^{\left( s\right) } &=&\left( \widetilde{B}_{1L}^{\left(
s\right) },\widetilde{B}_{2L}^{\left( s\right) }\right) \sim \left( \mathbf{2%
},1\right) ,\hspace{0.2cm}\widetilde{B}_{R}^{\left( s\right) }=\left( 
\widetilde{B}_{1R}^{\left( s\right) },\widetilde{B}_{2R}^{\left( s\right)
}\right) \sim \left( \mathbf{2},-1\right) ,  \notag \\
E_{L}^{\left( s\right) } &=&\left( E_{1L}^{\left( s\right) },E_{2L}^{\left(
s\right) }\right) \sim \left( \mathbf{2},1\right) ,\hspace{0.2cm}%
E_{R}^{\left( s\right) }=\left( E_{1R}^{\left( s\right) },E_{2R}^{\left(
s\right) }\right) \sim \left( \mathbf{2},1\right) ,  \notag \\
\nu _{sR} &=&\left( \mathbf{1}^{\prime },-1\right) ,\hspace{5mm}s=1,2,
\end{eqnarray}%
Here $\mathbf{1}$ and $\mathbf{1^{\prime }}$ are the trivial and nontrivial $S_{3}$ singlets, respectively.

\begin{figure*}[tbh]
\begin{center}
\includegraphics[scale=0.7]{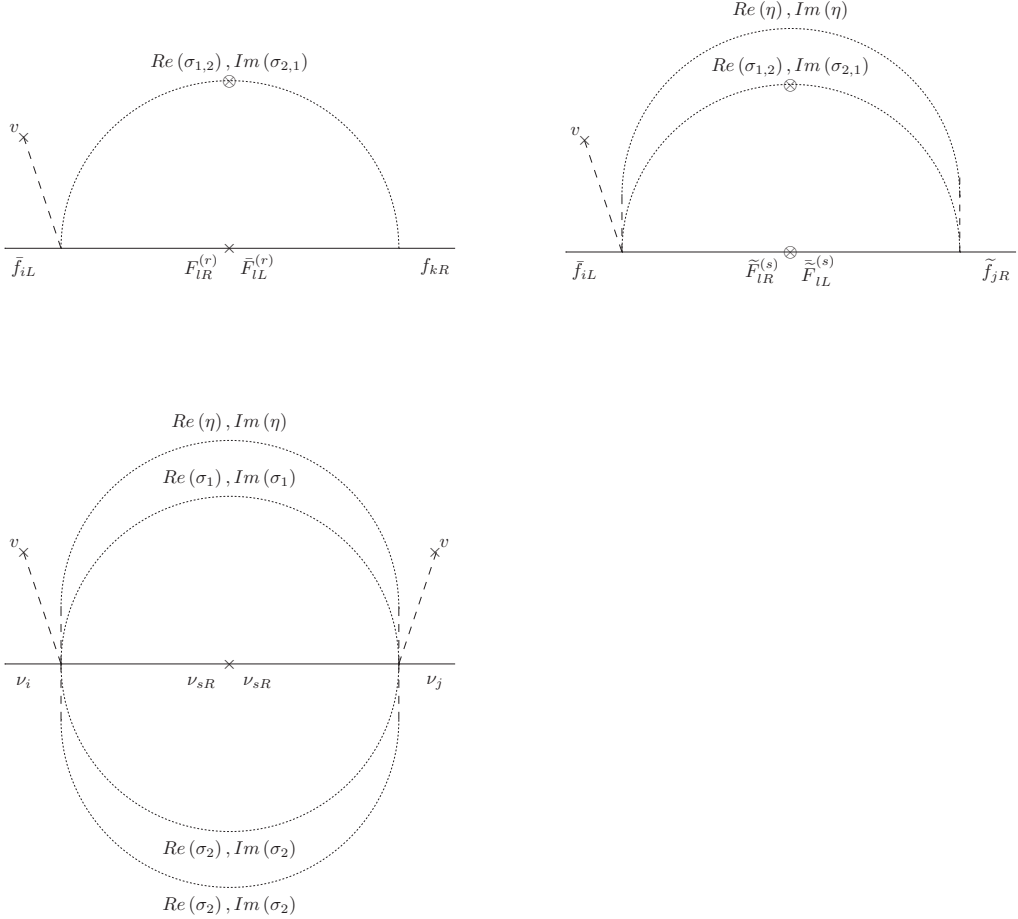}
\end{center}
\vspace{-5.0cm}
\caption{Loop Feynman diagrams contributing to the
fermion mass matrices. Here $f_{iL}=u_{iL},d_{iL},e_{iL}$ ($i=1,2,3$), 
$f_{R}=u_{2R},d_{3R},l_{2R},l_{3R}$, 
$\widetilde{f}_{R}=u_{1R},d_{1R},d_{2R},l_{1R}$, $l=1,2$. 
The EW singlet charged exotic
fermions, see (\ref{eq:Assign-1}) are denoted by $F_{lR}^{(r)}$, $F_{lL}^{(r)}$, $\widetilde{F}_{lR}^{(s)}$ and $\widetilde{F}_{lL}^{(s)}$. It is implied $r=1$ for quarks, $r,k=1,2$ for charged leptons, $s,j=1$ for up-type quarks and charged leptons, whereas $s,j=1,2$ for down-type quarks, $s=1,2$ and $j=1,2,3$ for neutrinos.}
\label{Oneloopdiagrams}
\end{figure*}

With this field content, the relevant quark, charged lepton and neutrino
Yukawa terms invariant under the symmetry (\ref{eq:symm-1}) 
take the form:
\begin{eqnarray}
-\mathcal{L}_{\text{Y}}^{\left( U\right) } &=&\sum_{j=1}^{3}y_{j}^{\left(
u\right) }\overline{q}_{jL}\widetilde{\phi }\left( \widetilde{T}_{R}\sigma
\right) _{\mathbf{1}}\frac{\eta }{\Lambda ^{2}} +x^{\left( u\right) }\left( \overline{\widetilde{T}}_{L}\sigma \right) _{%
\mathbf{1}^{\prime }}u_{1R}\frac{\eta }{\Lambda }  \notag \\
&&+\sum_{j=1}^{3}z_{j}^{\left( u\right) }\overline{q}_{jL}\widetilde{\phi }%
\left( T_{R}\sigma \right) _{\mathbf{1}}\frac{1}{\Lambda }+w^{\left( u\right) }\left( \overline{T}_{L}\sigma \right) _{\mathbf{1}%
^{\prime }}u_{2R}  \notag \\
&&+\sum_{j=1}^{3}y_{j3}^{\left( u\right) }\overline{q}_{jL}\widetilde{\phi }%
u_{3R}+y_{T}\left( \overline{T}_{L}T_{R}\right) _{\mathbf{2}}\sigma +h.c.  
\label{lyu}
\end{eqnarray}%
\begin{eqnarray}
-\mathcal{L}_{\text{Y}}^{\left( D\right) }
&=&\sum_{j=1}^{3}\sum_{s=1}^{2}y_{js}^{\left( d\right) }\overline{q}%
_{jL}\phi \left( \widetilde{B}_{R}^{\left( s\right) }\sigma \right) _{%
\mathbf{1}}\frac{\eta }{\Lambda ^{2}}+\sum_{s=1}^{2}\sum_{k=1}^{2}x_{sk}^{\left( d\right) }\left( \overline{%
\widetilde{B}}_{L}^{\left( s\right) }\sigma \right) _{\mathbf{1}^{\prime
}}d_{kR}\frac{\eta }{\Lambda }  \notag \\
&&+\sum_{j=1}^{3}z_{j}^{\left( d\right) }\overline{q}_{jL}\phi \left(
B_{R}\sigma \right) _{\mathbf{1}}\frac{1}{\Lambda }+w^{\left( d\right) }\left( \overline{B}_{L}\sigma \right) _{\mathbf{1}%
^{\prime }}d_{3R}+y_{B}\overline{B}_{L}B_{R}\sigma +h.c.  \label{lyd}
\end{eqnarray}%
\begin{eqnarray}
\label{eq:LeptYuk}
-\mathcal{L}_{\text{Y}}^{\left( l\right) } &=&\sum_{j=1}^{3}y_{j}^{\left(
l\right) }\overline{l}_{jL}\phi \left( \widetilde{E}_{R}\sigma \right) _{%
\mathbf{1}}\frac{\eta }{\Lambda ^{2}} +x_{1}^{\left( l\right) }\left( \overline{\widetilde{E}}_{L}\sigma \right)
_{\mathbf{1}^{\prime }}l_{1R}\frac{\eta }{\Lambda }  \notag \\
&&+\sum_{j=1}^{3}\sum_{s=1}^{2}y_{js}^{\left( l\right) }\overline{l}%
_{jL}\phi \left( E_{R}^{\left( s\right) }\sigma \right) _{\mathbf{1}}\frac{1%
}{\Lambda } +\sum_{k=1}^{2}\sum_{s=1}^{2}x_{ks}^{\left( l\right) }\left( \overline{E}%
_{L}^{\left( s\right) }\sigma \right) _{\mathbf{1}^{\prime }}l_{kR}  \notag
\\
&&+\sum_{s=1}^{2}y_{s}^{\left( E\right) }\left( \overline{E}_{L}^{\left(
s\right) }E_{R}^{\left( s\right) }\right) _{\mathbf{2}}\sigma+h.c. \label{lyl}, \\[3mm]
-\mathcal{L}_{\text{Y}}^{\left( \nu \right) }
&=&\sum_{j=1}^{3}\sum_{s=1}^{2}y_{js}^{\left( \nu \right) }\overline{l}_{jL}%
\widetilde{\phi }\nu _{sR}\frac{\left[ \sigma \left( \sigma \sigma \right) _{%
\mathbf{2}}\right] _{\mathbf{1}^{\prime }}\eta }{\Lambda ^{4}}+\sum_{s=1}^{2}\ m_{s} \bar{\nu}_{sR} \nu^{C}_{s R} +h.c.  \label{Lynu}  
\end{eqnarray}
%
%
%
We want that after the spontaneous breaking of the electroweak symmetry 
the above-given Yukawa interactions generate the SM fermion masses
according to (\ref{eq:MassHierarch-1})-(\ref{eq:MassHierarch-4}). This happens if we
introduce soft $Z_{2}$ breaking terms in the sector of the electroweak
singlet fermions 
\begin{eqnarray}
\mathcal{L}_{soft}^{F}=\widetilde{m}_{T}\left( \overline{\widetilde{T}}%
_{L}\widetilde{T}_{R}\right) _{\mathbf{1}}+\sum_{s=1}^{2}\widetilde{m}%
_{B}^{\left( s\right) }\left( \overline{\widetilde{B}}_{L}^{\left( s\right) }%
\widetilde{B}_{R}^{\left( s\right) }\right) _{\mathbf{1}}
+\widetilde{m}_{E}\left( \overline{\widetilde{E}}_{L}\widetilde{E}_{R}\right) _{\mathbf{1}}+ h.c.\,,\label{eq:Soft-1}
\end{eqnarray}
%
%
%
as well as soft $S_{3}$ breaking in the electroweak singlet scalar
sector 
\begin{eqnarray}\label{eq:Soft-2}
\mathcal{L}_{soft}^{\sigma }=\mu _{12}^{2}\sigma _{1}\sigma _{2}
\end{eqnarray}
for the $S_{3}$ scalar doublet $\sigma =\left( \sigma _{1},\sigma_{2}\right) $. 
From the interactions (\ref{lyu})-(\ref{Lynu}) there emerge 1-, 2- and 4-loop diagrams shown in Fig.~\ref{Oneloopdiagrams}.
They implement the loop hierarchical pattern of 
the SM fermion mass matrix entries (\ref{eq:MassHierarch-2})-(\ref{eq:MassHierarch-4}).
The top-quark entry $\kappa_{j3}$, according  to the field assignments in Eqs.~(\ref{eq:Assign-1}), 
is generated at tree-level as declared in (\ref{eq:MassHierarch-1}). 


Let us comment on the model setup (\ref{eq:symm-1})-(\ref{eq:Assign-1}). In its elaboration we were 
guided 
by minimality arguments,
compatible with the hierarchical structure (\ref{eq:MassHierarch-1})-(\ref{eq:MassHierarch-4}).  The selection of 
the discrete group (\ref{eq:Assign-1}) is motivated by the following reasons. The $S_{3}$ is the smallest non-abelian group having a doublet irreducible
representation, necessary to set up a minimally non-trivial structure of Yukawa interactions (\ref{lyu})-(\ref{Lynu}), leading to 
(\ref{eq:MassHierarch-1})-(\ref{eq:MassHierarch-4}). 
We also need a preserved $Z_{2}$ symmetry to separate
the exotic $\widetilde{F}_{R}^{\left( s\right) }$ and the SM $\widetilde{f}_{R}=u_{1R},d_{1R},d_{2R},l_{1R}$ fermionic fields from the remaining fermions,
as done in the list of the assignments (\ref{eq:Assign-1}).  The diagrams in Fig.~\ref{Oneloopdiagrams} require the $S_{3}\times Z_{2}$ soft breaking 
terms shown in Eqs.~(\ref{eq:Soft-1})-(\ref{eq:Soft-2}). The corresponding mass-insertions are marked in the diagrams with the $\otimes$ sign on the $F$- and $\sigma$-lines.
In our model we trade renormalizability for minimality, in the sense that within the setup (\ref{eq:symm-1})-(\ref{eq:Assign-1}) we need to have non-renormalizable operators present in 
the Lagrangians (\ref{lyu})-(\ref{Lynu}), with some characteristic scale $\Lambda$.  We assume this scale to be common for all the operators.  These operators are crucial for the construction of the diagrams in Fig.~\ref{Oneloopdiagrams}, which realize in conventional terminology the 1-, 2- and 4-loop radiative seesaw mechanisms.  The presence of the scale $\Lambda$ implies that there are heavy degrees of freedom of mass $M \gtrsim \Lambda$, corresponding to an ultraviolet renormalizable completion of our model. 
In Section \ref{rm} we discuss some possible renormalizable models leading below the cutoff scale 
$\Lambda$ to the effective operators of our model.
Let us estimate this scale from the observable fermion masses, taking as an example the 1-loop $b$-quark and 
2-loop strange quark masses: $m_{s}$ and $m_{b}$, respectively.
Estimating the first and second diagrams in Fig.~\ref{Oneloopdiagrams}, 
we get an order of magnitude value
%
\begin{eqnarray}
\label{eq:Lambda-Estimate-b} 
m_b &\sim& \frac{y^2_b}{16\pi^2}f_1\frac{v}{\Lambda}\frac{\mu_{12}}{M}\mu_{12},\\
\label{eq:Lambda-Estimate-s}
 m_s &\sim& \frac{y^2_s}{\left(16\pi^2\right)^2}f_2\frac{v}{M}\frac{\mu^3_{12}}{\Lambda^3}\mu_{12},
\end{eqnarray}
where $f_{1,2}$ 
are functions of the masses $M$ of the particles running inside the loops and 
$y^2_{b,s}$
are 
the products of two Yukawa couplings in the vertices of the diagrams in Fig.~\ref{Oneloopdiagrams}.
Assuming $y^2_b f_{1}\sim y^2_s f_{2} \sim 1$ 
and $\mu_{12}\sim M$, we find a rough estimate 
\begin{eqnarray}\label{eq:Lambda-EstimateVal}
&&\Lambda\sim 10v\sim 2.5{\rm TeV}
\end{eqnarray}
for the correct order of magnitude of $m_{b}$ and $m_{s}$.

\section{Model Phenomenology} 
Let us recall that our goal is to explain the hierarchy  (\ref{eq:MassHierarch-1})-(\ref{eq:MassHierarch-4}),
without pretending to predict the values of the quark and lepton masses and mixings. Nevertheless, 
resorting to reasonable assumptions about the model parameters and using (\ref{eq:Lambda-EstimateVal}), 
we are able to give several predictions at least at an order of magnitude accuracy. 


In the neutrino sector,
from the 4-loop diagrams in Fig.~\ref{Oneloopdiagrams}, it follows that the light active neutrino
mass matrix takes the form%
\begin{eqnarray}\label{eq:NuMassMatr-1}
M_{\nu }&=&\frac{\mu _{\eta }^{2}\mu _{\sigma }^{6}v}{\left( 16\pi
^{2}\right) ^{4}\Lambda ^{8}} 
\nonumber
\left(  
\begin{array}{ccc}
\beta _{1}^{2}+\gamma _{1}^{2} & \beta _{1}\beta _{2}+\gamma _{1}\gamma _{2}
& \beta _{1}\beta _{3}+\gamma _{1}\gamma _{3} \\  
\beta _{1}\beta _{2}+\gamma _{1}\gamma _{2} & \beta _{2}^{2}+\gamma _{2}^{2}
& \beta _{2}\beta _{3}+\gamma _{2}\gamma _{3} \\  
\beta _{1}\beta _{3}+\gamma _{1}\gamma _{3} & \beta _{2}\beta _{3}+\gamma
_{2}\gamma _{3} & \beta _{3}^{2}+\gamma _{3}^{2}%
\end{array}
\right) ,  \\[3mm]
\label{eq:NuMassMatr-2}
\beta _{s} &=&y_{s1}^{\left( \nu \right) }\frac{v}{m_{1}}f_{1}^{\left( \nu
\right) },\hspace{3mm}\gamma _{s}=y_{s2}^{\left( \nu \right) }\frac{v}{m_{2}}
f_{2}^{\left( \nu \right) },\hspace{3mm}s=1,2.
\end{eqnarray}
where $m_{1}$ and $m_{2}$ are the heavy right handed Majorana neutrino, 
$\nu_{sR}$,
masses and $f_{1}^{\left( \nu \right) }$, $f_{2}^{\left( \nu \right) }$ are
functions which depend on the masses of the particles running inside the 4-loop diagrams.  The structure of the mass matrix in Eq.~(\ref{eq:NuMassMatr-1}) is so that 
$\det \left( M_{\nu }\right) $ $=0$. Thus, there is at least one massless neutrino. It can be checked directly that the number of the massless states 
is $3-n_{R}$, where $n_{R}$  is the number of massive right handed Majorana neutrinos $\nu_{sR}$. In order to be compatible with the neutrino oscillation data, we need 
at least two light massive active neutrinos. That is why we introduced in the model (\ref{eq:Assign-1}) 
two massive right-handed neutrinos $\nu_{1R}, \nu_{2R}$, which is the minimal number necessary for this purpose, both for  normal and inverted neutrino mass hierarchy.  

For similar reasons we introduced in the model (\ref{eq:Assign-1}) a minimal number of the exotic fermions $T, \tilde{T}$, $B, \tilde{B}$ and $E, \tilde{E}$,  necessary so that no massless charged SM-fermions would appear in the model.

The typical mass of the two light active neutrinos can be estimated from Eqs.~(\ref{eq:NuMassMatr-1}), (\ref{eq:NuMassMatr-2}). It is roughly
%
\begin{eqnarray}\label{eq:NuMass-1}
&&m_{\nu}\sim\frac{\left(y^{(\nu)}\right)^{2}}{\left(16\pi^2\right)^4}f^{(\nu)} \frac{v}{m_{s}}\frac{\mu^2_{\eta}\mu^6_{\sigma}}{\Lambda^8}v. 
\end{eqnarray}
%
%
Assuming  $\left(y^{(\nu)}\right)^{2}\cdot f^{(\nu)} \sim 1$, $\mu_{\eta}\sim\mu_{\sigma}\sim m_{s}\sim \alpha\cdot \Lambda$
and taking $\Lambda = 2.5$TeV from the quark sector (\ref{eq:Lambda-EstimateVal})
we find for $\alpha \sim 1$ the light neutrino mass scale $m_{\nu}\sim 1$eV, which is too heavy. However in our model all the particles are lighter than the cutoff scale $\Lambda$. Assuming, for instance,  
$\alpha = 0.3$ we arrive at the correct neutrino mass scale 
\mbox{$m_{\nu}\sim 50$ meV.}


Let us survey the possible dark matter (DM) particle candidates in our model.
Due to the preserved $S_{3}\times Z_{2}$ symmetry, 
this role could be assigned either to the right handed Majorana neutrinos $\nu_{sR}$ or to the lightest of 
the scalar fields $\eta $, $Re(\sigma _{s})$ and $Im(\sigma _{s})$ ($s=1,2$). Here we analyse the case in which the SM singlet $Z_{2}$-odd scalar particle $\eta$ is lighter than the $\sigma_1$ and $\sigma_2$ scalars.  
In this mass range the $\eta$ is stable.  In fact, the only possible decay modes of 
$\eta$ are
%
\begin{eqnarray}
\eta&\to&\sigma_{1,2}\widetilde{T}_{{2L,1L}}u_{1R}, \, \sigma_{1,2}\widetilde{T}_{{1R,2R}}u_{iL}, \  \sigma_{1,2}\widetilde{B}^{(s)}_{{2L,1L}}d_{kR}, \  \sigma_{1,2}\widetilde{B}^{(s)}_{{1R,2R}}d_{iL}, \  \sigma_{1,2}\widetilde{E}_{{2L,1L}}l_{1R},\notag\\
&&\sigma_{1,2}\widetilde{E}_{{1R,2R}}e_{iL}, \  \sigma_{1}2\sigma_{2}\nu_{iL}\nu_{sR}\label{eq:Eta-Decs-1}
\end{eqnarray}
with $s,k = 1,2$ and $i=1,2,3$. These decays may arise from the first and second terms of the charged fermion Yukawa interactions of Eqs.~(\ref{lyu}),  (\ref{lyd}) and (\ref{lyl}) as well as from the first term of the neutrino Yukawa interaction of Eq.~(\ref{Lynu}), respectively. For $\eta$ lighter than $\sigma_{1,2}$, the decays (\ref{eq:Eta-Decs-1})  are kinematically forbidden, and as a result, the $\eta$ is stable, as necessary for a DM particle candidate.
%
%
Let us estimate its relic density according to 
(c.f. Ref.~\cite{Olive:2016xmw})
%
\begin{eqnarray}\label{eq:relic-1}
&&\Omega h^{2}=\frac{0.1pb}{\left\langle \sigma v\right\rangle },\,\hspace{1cm}\left\langle \sigma v\right\rangle=\frac{A}{n_{eq}^{2}}\,,
\end{eqnarray}
%
where $\left\langle \sigma v\right\rangle $ is the thermally averaged annihilation cross-section, $A$ is the total annihilation rate per unit volume at temperature $T$ and  $n_{eq}$ is the equilibrium value of the particle density, which are given by \cite{Edsjo:1997bg}
\begin{eqnarray}
A&=&\frac{T}{32\pi ^{4}}\dint\limits_{4m_{\eta }^{2}}^{\infty
}\dsum\limits_{p=W,Z,t,b,h}g_{p}^{2}\frac{s\sqrt{s-4m_{\eta }^{2}}}{2}
v_{rel}\sigma \left( \eta \eta \rightarrow p\overline{p}\right)
K_{1}\left( \frac{\sqrt{s}}{T}\right) ds,\notag \\
n_{eq} &=&\frac{T}{2\pi ^{2}}\dsum\limits_{p=W,Z,t,b,h}g_{p}m_{\eta
}^{2}K_{2}\left( \frac{m_{\eta }}{T}\right),
\end{eqnarray}
with $K_{1}$ and $K_{2}$ being the modified Bessel functions of the second kind order 1 and 2, respectively \cite{Edsjo:1997bg}. For the relic density calculation, we take $T=m_{\eta }/20$ as in Ref. \cite{Edsjo:1997bg}, which corresponds to a typical freeze-out temperature. 
We assume that our DM candidate $\eta$ annihilates mainly into $WW$, $ZZ$, $t\overline{t}$, $b\overline{b}$ and $hh$,
with annihilation cross sections 
\cite{Bhattacharya:2016ysw}:
\begin{eqnarray}
v_{rel}\sigma \left( \eta \eta \rightarrow WW\right)  &=&\frac{\lambda
_{h^{2}\eta ^{2}}^{2}}{8\pi }\frac{s\left( 1+\frac{12m_{W}^{4}}{s^{2}}-\frac{%
4m_{W}^{2}}{s}\right) }{\left( s-m_{h}^{2}\right)^{2} +m_{h}^{2}\Gamma _{h}^{2}}
\sqrt{1-\frac{4m_{W}^{2}}{s}},  \notag \\
v_{rel}\sigma \left( \eta \eta \rightarrow ZZ\right)  &=&\frac{\lambda
_{h^{2}\eta ^{2}}^{2}}{16\pi }\frac{s\left( 1+\frac{12m_{Z}^{4}}{s^{2}}-%
\frac{4m_{Z}^{2}}{s}\right) }{\left( s-m_{h}^{2}\right)^{2}+m_{h}^{2}\Gamma
_{h}^{2}}  
\sqrt{1-\frac{4m_{Z}^{2}}{s}},  \notag \\
v_{rel}\sigma \left( \eta \eta \rightarrow q\overline{q}\right)  &=&\frac{%
N_{c}\lambda _{h^{2}\eta ^{2}}^{2}m_{q}^{2}}{4\pi}\frac{\sqrt{\left(1-\frac{%
4m_{f}^{2}}{s}\right)^{3}}}{\left( s-m_{h}^{2}\right)^{2} +m_{h}^{2}\Gamma _{h}^{2}}, 
\notag \\
v_{rel}\sigma \left( \eta \eta \rightarrow hh\right)  &=&\frac{\lambda
_{h^{2}\eta ^{2}}^{2}}{16\pi s}\left( 1+\frac{3m_{h}^{2}}{s-m_{h}^{2}}-\frac{%
4\lambda _{h^{2}\eta ^{2}}v^{2}}{s-2m_{h}^{2}}\right) ^{2} 
\sqrt{1-\frac{4m_{h}^{2}}{s}},
\end{eqnarray}
where $\sqrt{s}$ is the centre-of-mass energy, $N_{c}=3$ is the color factor,
$m_{h}=125.7$ GeV and $\Gamma _{h}=4.1$ MeV are the SM Higgs boson $h$ mass and
its total decay width, respectively.

\begin{figure}[t]
\center
\vspace{0.8cm} \subfigure{\includegraphics[width=0.7%
\textwidth]{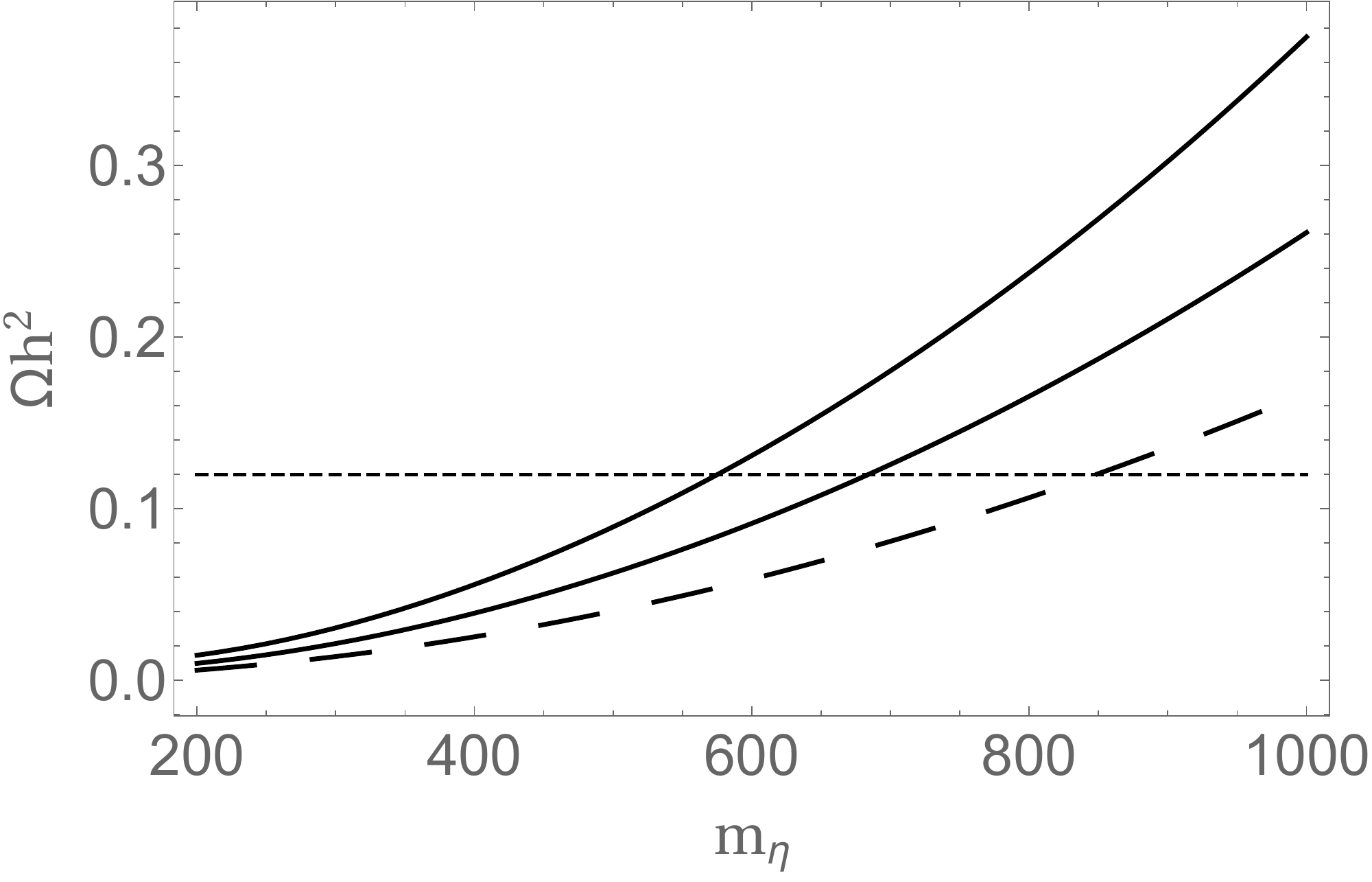}}
\caption{Relic density $\Omega h^2$, as a function of the $Z_2$ odd scalar
mass $m_{\protect\eta}$, for several values of the quartic scalar coupling $%
\protect\lambda_{h^2\protect\eta^2}$. The curves from top to bottom
correspond to $\protect\lambda_{h^2\protect\eta^2}=1,1.2,1.5$,
respectively. The horizontal line shows the observed value $\Omega h^2=0.1198$ \cite{Ade:2015xua} for the relic density.}
\label{Figure1}
\end{figure}

Fig.~\ref{Figure1} displays the Relic density $\Omega h^{2}$ as a
function of the $Z_{2}$-odd scalar mass $m_{\eta }$, for several values of
the quartic scalar coupling $\lambda _{h^{2}\eta ^{2}}$.
The curves from top to bottom correspond to $\protect\lambda_{h^2\protect\eta^2}$ =1, 1.2 and 1.5, respectively. The horizontal line shows the observed value $\Omega h^2=0.1198$.
As can be seen, the Relic density is an increasing function of  $m_{\eta }$ and a decreasing function of 
$\protect\lambda_{h^2\protect\eta^2}$.   
In our model we expect a typical mass scale for  all the non-SM particles -- the $\eta$-DM candidate, in particular, -- to be  $m_{\rm non-SM}\sim m_{\eta} \sim \alpha \cdot\Lambda \sim 750$~GeV.  This is hinted, as motivated below Eq.~(\ref{eq:NuMass-1}), by the light neutrino mass scale \mbox{$m_{\nu}\sim 50$ meV}.  For this DM particle mass, as seen from Fig.~\ref{Figure1}, the quartic coupling  is $\protect\lambda_{h^2\protect\eta^2}/(4 \pi) \leq 1$, which corresponds to the perturbative regime of the model.


Finally we briefly comment about the possible implications of our model for LHC searches.
Curiously, the typical mass scale of the non-SM particles is in our model about 750~GeV, resembling 
the diphoton excess reported by the ATLAS and CMS collaborations in 2015. Although it has been recently declared
to be a statistical fluctuation, the scalars in this mass ballpark appear naturally in various models.  

The exotic $T_{n}$, $\widetilde{T}_{n}$, $B_{n}$, $\widetilde{B}_{n}^{\left( s\right) }$, ($n,s=1,2$)  quarks and the exotic charged leptons $E_{n}^{\left( s\right) }$, $\widetilde{E}_{n}$ ($n,s=1,2$) are also assumed to be around this mass scale. Therefore they can be produced in pairs at the LHC, via both the gluon fusion and the Drell-Yan mechanism in the case of the exotic quarks, and only via the Drell-Yan in the case of the charged exotic leptons. 
From the charged fermion Yukawa interactions (\ref{eq:LeptYuk}), it follows that the exotic charged fermions can be searched at the LHC through their decays into SM charged fermions and a single or a couple of SM scalar singlets. Thus the signal would be an excess of events with respect to the SM background in the dijet and opposite sign dilepton final states. A more detailed analysis will be done elsewhere. 

\section{UV completions of the Model}
\label{rm}
Finally we comment on  the possible  ultraviolet (UV) origin of the non-renormalizable Yukawa terms in Eqs.~(\ref{lyu})-(\ref{Lynu}).  Let us list them together
%
\begin{eqnarray}\label{NRchargedfermions}
&&\overline{f}_{L}H\left( F_{R}\, \sigma \right) _{\mathbf{1}}
\frac{1}{\Lambda },\hspace{1cm}
\overline{f}_{L}H\left( \widetilde{F}_{R}\, \sigma \right) _{\mathbf{1}}\frac{\eta}{\Lambda ^{2}},\hspace{1cm}
\left(\overline{\widetilde{F}}_{L}\, \sigma\right) _{\mathbf{1}^{\prime }}
\widetilde{f}_{R}\frac{\eta}{\Lambda},
%
\\
\label{NRneutrinos}
&&\overline{l}_{L}%
\widetilde{\phi }\nu _{R}\frac{\left[ \sigma \left( \sigma \sigma \right) _{\mathbf{2}}\right] _{\mathbf{1}^{\prime }}\eta }{\Lambda ^{4}}\, .
\end{eqnarray}
Here we  suppressed all the super- and subscripts of the fields, unessential for our discussion. We denoted with $F_{L,R}, \widetilde{F}_{L,R}$ the exotic electroweak singlet quarks 
$T_{L,R}, B_{L,R}, \widetilde{T}_{L,R}, \widetilde{B}_{L,R}$ and charged leptons 
$E_{L,R}, \widetilde{E}_{L,R}$ introduced in Eqs.~(\ref{eq:Assign-1}), and
the SM fermions with $f_{L}=q_{jL},l_{jL}$; $l_{L} = l_{jL}$ ($j=1,2,3$); 
 and 
$f_{R}=u_{2R},d_{3R},l_{2R},l_{3R}$ and $\widetilde{f}_{R}=u_{1R},d_{1R},d_{2R},l_{1R}$.
In the first two terms it is implied that $H\equiv \widetilde{\phi}$ for $F=T$ and $\widetilde{F}=\widetilde{T}$, while $H\equiv \phi$ for $F=B, E$ and $\widetilde{F} = \widetilde{B}, \widetilde{E}$. 

%
%

These four non-renormalizable Yukawa terms of Eqs. (\ref{NRchargedfermions}) and (\ref{NRneutrinos}) can be generated at low energies by the Feynman diagrams shown in Fig.~\ref{DiagramsOperators}, after integrating out the heavy scalar fields $\xi$, $\Phi$, $\chi$, $\varphi$, $\rho$ with characteristic masses of the order of our model cutoff scale $\Lambda$. Their assignment to the model symmetry group (\ref{eq:symm-1}) is dictated by the requirement that renormalizable interactions in the vertices of these diagrams be singlets with respect to this group. Thus we find the  
$S_{3}\otimes Z_{2}$ assignments for these heavy scalars:
\begin{figure*}[tbh]
\begin{center}
\includegraphics[scale=0.7]{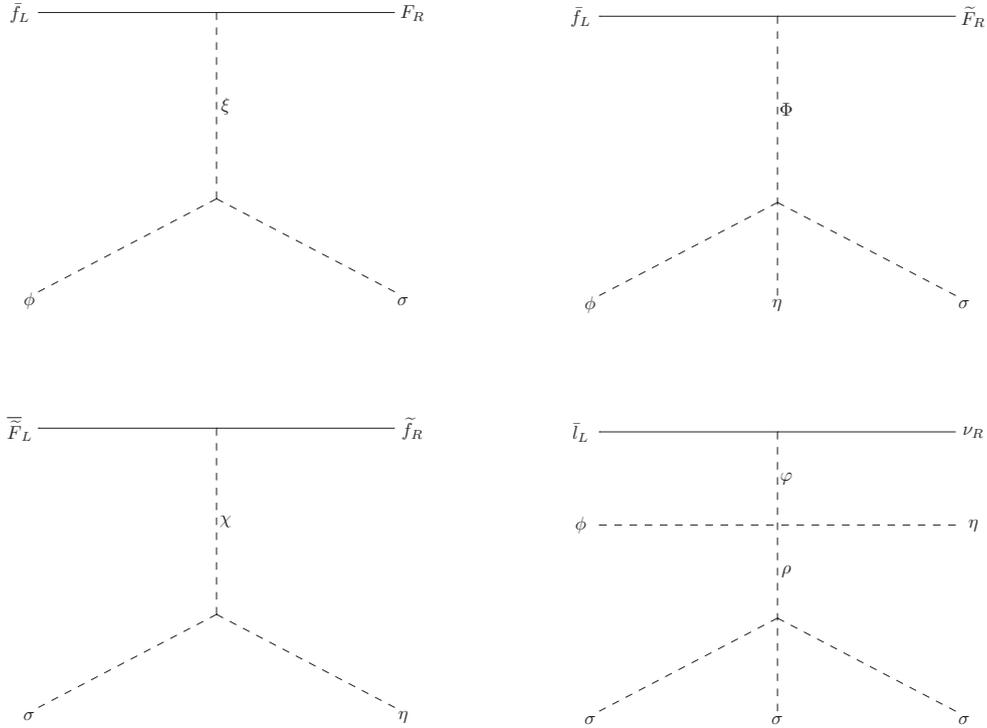}
\end{center}
\vspace{-8.0cm}
\caption{Feynman diagrams that induce the non-renormalizable operators of Eqs. \ref{NRchargedfermions} and \ref{NRneutrinos}. 
}
\label{DiagramsOperators}
\end{figure*}
%
\begin{eqnarray}
\xi&=&\left(\xi_1,\xi_2\right)\sim\left( \mathbf{2},1\right),\hspace{1cm}\Phi=\left(\Phi_1,\Phi_2\right)\sim\left( \mathbf{%
2},-1\right),\hspace{1cm}\chi=\left(\chi_1,\chi_2\right)\sim\left( \mathbf{%
2},-1\right)\notag\\
\varphi&=&\left( \mathbf{1}^{\prime },-1\right),\hspace{1cm}\rho=\left( \mathbf{1}^{\prime },1\right)
\end{eqnarray}
where $\xi_i$, $\Phi_i$ $(i=1,2)$ and $\varphi$ are $SU_{2L}$ doublets with hypercharge of $\frac1{2}$ (as the SM Higgs doublet $\phi$), whereas $\chi_i$ $(i=1,2)$ and $\rho$ are SM singlets with zero hypercharge.

In this particular UV completion the non-renormalizable operators  (\ref{NRchargedfermions}),  (\ref{NRneutrinos}) are replaced with the renormalizable interactions
\begin{eqnarray}
\mathcal{L}_{UV} &\propto &
z_{\zeta }\,\overline{f}_{L}\left(\zeta\, F_{R}\right) _{\mathbf{1}}+
z_{\Xi }\overline{f}_{L}\left(\Xi \,  \widetilde{F}_{R}\right) _{\mathbf{1}}+
z_{\chi }\left(\overline{\widetilde{F}}_{L}\chi\right) _{\mathbf{1}^{\prime }}\widetilde{f}_{R}+
z_{\varphi }\overline{l}_{L}\widetilde{\varphi }\nu _{R}\label{UV} \\
&&+\mu _{\phi \xi \sigma }\left( \phi \cdot \xi ^{\dagger }\right) \sigma
+\lambda _{\phi \Phi \sigma \eta }\left( \phi \cdot \Phi ^{\dagger }\right)
\sigma \eta +\mu _{\sigma \eta \chi }\sigma \eta \chi +\lambda _{\rho \sigma
^{3}}\rho \sigma ^{3}+\lambda _{\phi \varphi \eta \rho }\left( \phi \varphi
^{\dagger }\right) \eta \rho +.... \notag 
\end{eqnarray}
corresponding to the vertices of the diagrams in Fig.~\ref{DiagramsOperators}.
Here $\zeta\equiv \widetilde{\xi}$, $\Xi\equiv \widetilde{\Phi}$ for $F_{R} = T_{R}$ and 
$\widetilde{F}_{R} = \widetilde{T}_{R}$
while 
$\zeta\equiv \xi$, $\Xi\equiv \Phi$ for $F_{R} = B_{R}, E_{R}$ and 
$\widetilde{F}_{R} = \widetilde{B}_{R}, \widetilde{E}_{R}$. The  Yukawa and scalar  self-interaction couplings we denoted with $z$ and $\mu, \lambda$, respectively.
This is just one of many possible UV completions of our effective model.  In the present paper we do not intend to list all of them, although this is a quite straightforward group theory exercise. Going upwards in the energy scale with a particular UV model one may speculate on the unification to an extended gauge symmetry group relating
some parameters of the Lagrangian (\ref{lyu})-(\ref{Lynu}) and making the framework more predictive.  This extended symmetry should be spontaneously broken down to the group (\ref{eq:symm-1}) 
at a scale above $\Lambda$. This study is beyond the scope of the present paper.

\section{Conclusions.}
We have proposed the first model with the SM fermion mass hierarchy generated by the 
loops.   
We constructed a model setup as the minimal extension of the SM which allowed us to  
realize the radiative mechanism. The model does not pretend to explain the quark and lepton masses and mixing angles, but only the mass hierarchy. 
Nevertheless, through reasonable assumptions about the model parameters we were able to make several rough predictions with order of magnitude accuracy.
The model contains non-renormalizable operators with a cutoff scale $\Lambda$, which separates the dynamic particles with masses $\leq \Lambda$ 
from the heavy frozen degrees of freedom. We estimated this scale to be $\Lambda \sim 2.5$ TeV, from the 1- and 2-loop quark masses (\ref{eq:Lambda-Estimate-b}), (\ref{eq:Lambda-Estimate-s}).
In the neutrino sector our model predicts -- independently of the model parameter values -- one massless and two non-zero mass neutrinos: a mass spectrum compatible with the neutrino oscillation data. 
Analyzing the 4-loop neutrino mass (\ref{eq:NuMass-1}), we hinted that the mass scale of the non-SM particles of our model are of the order of 1~TeV. 
Due to the discrete symmetries, our model possesses DM particle candidates. We found that one of them,  the SM-singlet scalar $\eta$ lighter than the other non-SM scalars, could be a viable DM particle. We also commented on 
the possible implications of the exotic colored fermions for LHC searches. Finally we discussed one of the ultraviolet completions of our effective model.
\label{conclusions}

\subsection*{Acknowledgements}
This work was partially supported by Fondecyt (Chile), Grants 
No.~1150792, No.~1140390, No.~3150472 and by 
CONICYT (Chile) Ring ACT1406 and CONICYT PIA/Basal FB0821.

\end{document}